\documentclass[pra,aps,twocolumn,showpacs,amsmath,amssymb]{revtex4}
\usepackage{graphicx}
\usepackage{dcolumn}
\usepackage{bm}
\usepackage{subfigure}
\usepackage{setspace}
\begin{document}
\title{Atom Fock state preparation by trap reduction}
\author{A. del Campo}
\email{adolfo.delcampo@ehu.es} \affiliation{Departamento de Qu\'\i mica-F\'\i sica, Universidad del Pa\'\i s Vasco, Apartado 644, 48080 Bilbao, Spain}
\author{J. G. Muga}
\email{jg.muga@ehu.es} \affiliation{Departamento de Qu\'\i mica-F\'\i sica, Universidad del Pa\'\i s Vasco, Apartado 644, 48080 Bilbao, Spain}

\def\d{{\rm d}}
\def\la{\langle}
\def\ra{\rangle}
\def\om{\omega}
\def\Om{\Omega}
\def\vep{\varepsilon}
\def\wh{\widehat}
\def\tr{\rm{Tr}}
\def\da{\dagger}
\newcommand{\beq}{\begin{equation}}
\newcommand{\eeq}{\end{equation}}
\newcommand{\beqa}{\begin{eqnarray}}
\newcommand{\eeqa}{\end{eqnarray}}
\newcommand{\intf}{\int_{-\infty}^\infty}
\newcommand{\into}{\int_0^\infty}
\date{\today}
\begin{abstract}

We study the production of low atom number Fock states by reducing suddenly the potential trap in a 
1D strongly interacting (Tonks-Girardeau) gas. The fidelity of the Fock state preparation 
is characterized by the average and variance of the number of trapped atoms. 
Two different methods are considered: making the trap shallower
(atom culling [A. M. Dudarev {\it et al.}, Phys. Rev. Lett. {\bf 98}, 063001 (2007)],
also termed ``trap weakening'' here) 
and making the trap narrower (trap squeezing).
When used independently, the efficiency of both procedures is limited 
as a result of the truncation of the final state 
in momentum or position space with respect to the ideal atom number state. However, their combination provides 
a robust and efficient strategy to create ideal Fock states.

\end{abstract}
\maketitle
%
%

%
%
%
%

The generation of Fock states with a definite, controlled atomic number 
is a highly desirable objective 
both from fundamental and applied points of view. They may be useful  
for studying few particle interacting systems \cite{Phillips,tunn}, 
entanglement \cite{e},
or number and spin-squeezed atomic systems \cite{r1,r2}. 
Production of fotonic Fock states \cite{r3} 
and interferometric schemes for sub-shot-noise sensitivity approaching the Heisenberg limit \cite{Heislimit}
do also require input Fock states.

A necessary step towards this goal is the development of atom counting devices  
paving the way into 
quantum atom statistics \cite{atomcounting}. Indeed  
a technique with nearly unit efficiency has already been demonstrated \cite{CSMHPR05}. 
Moreover, the recently proposed method of atom culling by making the atom trap shallower (in the 
following also termed ``trap weakening'') \cite{CSMHPR05,DRN07} has achieved 
sub-Poissonian atom-number fluctuations for $60-300$ trapped atoms when adiabatic conditions were fulfilled, 
i.e., when the trap depth is varied slowly. 
(The reference case of Poissonian statistics 
is realized by the number of particles
in a small volume of a classical ideal gas.)
In this method the initial state 
is assumed to be a ground state for an unknown number of bosons, in general  
smaller than the maximum capacity of the initial trap (this is, the maximum number of particles that can be confined in the trap). This capacity depends on the trap characteristics and interatomic interaction. 
As the barrier height of the trap is slowly reduced and the maximum capacity is surpassed, 
the excess of atoms will leave the trap to produce eventually the  
Fock state corresponding to the maximum capacity of the final 
trap configuration. For a pictorial representation see Fig. 1 (upper panel). 
\begin{figure}[t]
\includegraphics[width=6cm,angle=0]{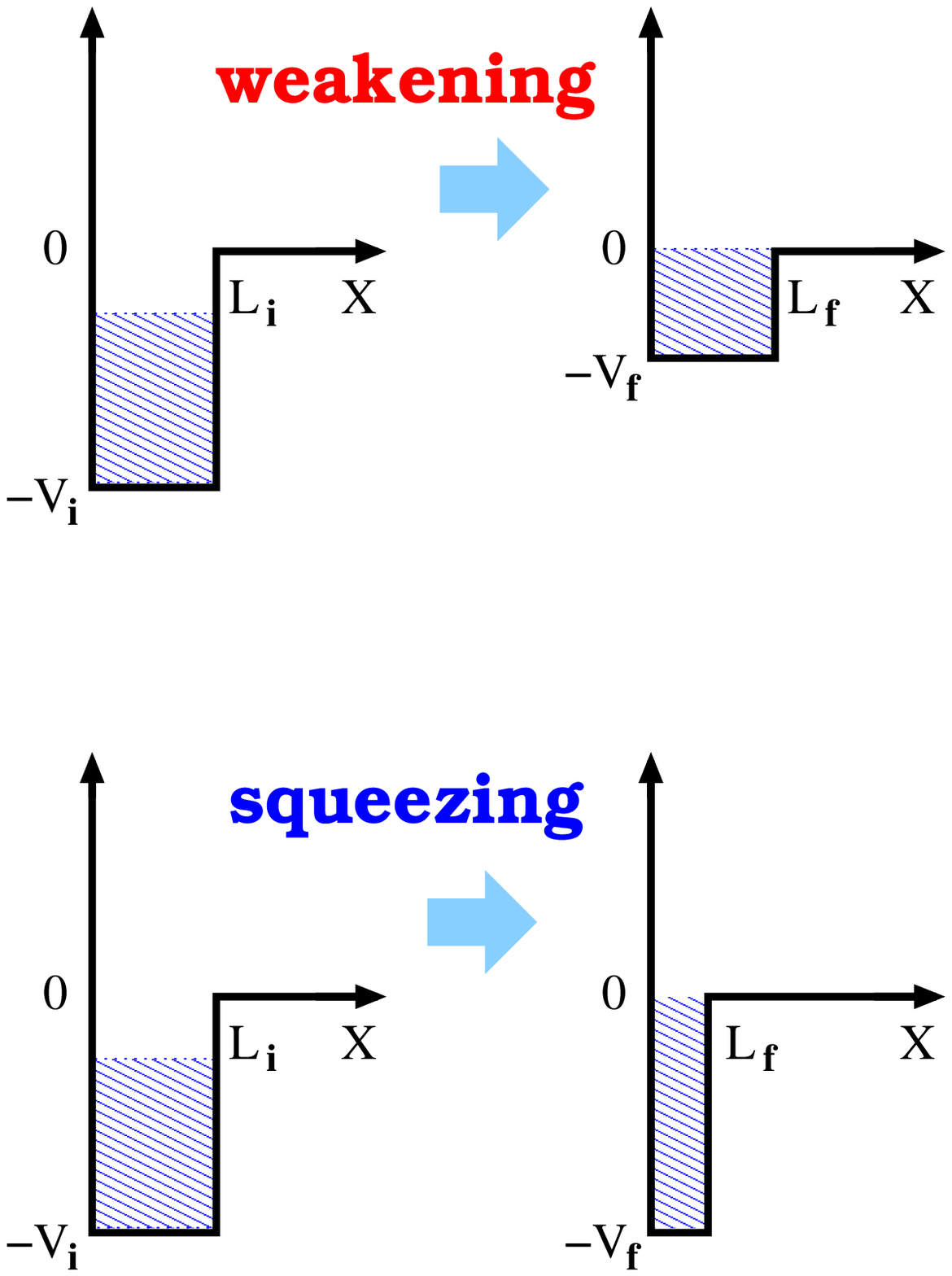}

\vspace{1.6cm}

\includegraphics[width=7cm,angle=0]{stat_fig1b.eps}
\caption{\label{pot}
(color online) Schematic potential change for trap weakening and squeezing, 
where $L_i$ and $L_f$ are the initial and final well widths, 
and $V_i$ and $V_f$ the initial and final trap depths, respectively.
Each solid line in the contour plot represents 
an isospectral family of traps at the threshold of a new bound state 
(for the initial trap $2mV_iL_{i}^{2}/\hbar^2=25\pi^2$, $C_f$ being the final trap capacity).
A transition between two given families is achieved 
reducing the trap capacity. The bound energy levels are pushed up and 
the higher ones cross the continuum threshold so that 
the excess of atoms escapes from the trap.
}
\end{figure}

A theoretical analysis has shown the basic properties of atom culling regarding 
the final average number 
as a function of the trap well depth and atom-atom interaction strength, covering the limits of the Tonks-Girardeau (TG) gas and the mean-field regime \cite{DRN07}: 
a weaker dependence on laser fluctuations of the height of the barrier that forms the trap 
will be favoured by strong interatomic interactions 
that separate the energy levels.
This motivates the present 
work, in which we focus on the strongly interacting 1D TG limit, optimal for atom culling,  
and examine the 
average number and fluctuations of the trapped atoms. 
Particular attention is paid to the sudden regime, 
corresponding to the ``worst case scenario'' of an abrupt change
from the initial to the final trap. 
We show that, even in this case, a state arbitrarily close to the ideal Fock state 
may be robustly produced by combining weakening and squeezing of
the trap, the two basic processes represented schematically 
in Fig. 1. 

\section{Atom statistics} 
Ultracold bosonic atoms in waveguides tight enough so that the transverse degrees of freedom are frozen out,
are well described by the Lieb-Liniger (LL) model \cite{LieLin63}.
%
In the strongly interacting limit \cite{Olshanii98,PSW00}
(for low densities and/or large one-dimensional scattering length) 
a LL gas tends to the TG gas,
which plays a distinguished role in atom statistics since its 
spatial antibunching has been predicted \cite{Kheruntsyan03}, and observed \cite{exp}.  
The system exhibits ``fermionization'' \cite{Girardeau60}, i.e., the 
TG gas and its ``dual'' system of spin-polarized ideal fermions behave similarly,
and share the same one-particle spatial density as well as any other local-correlation function, 
while differ on the non-local correlations.


The fermionic many-body ground state wavefunction of the dual system is built at time $t=0$ as a Slater determinant for $N_i$ 
particles,
$ \psi_{F}(x_{1},\dots,x_{N_i}) =\frac{1}{\sqrt{N_i!}}{\rm det}_{n,k=1}^{N_i}\varphi_{n}^i(x_{k})$,
where $\varphi_{n}^i(x)$ is the $n-$th eigenstate of the initial trap, whose time evolution will be denoted by $\varphi_{n}(x,t)$ whenever the external trap
is modified. 
The bosonic wave function, 
symmetric under permutation of particles, is obtained from $\psi_F$ by
the Fermi-Bose (FB) mapping \cite{Girardeau60,CS99} 
$
\psi(x_{1},\dots,x_{N_i})= \mathcal{A}(x_{1},\dots,x_{N_i})\psi_{F}(x_{1},\dots,x_{N_i})
$, 
where $\mathcal{A}=\prod_{1\leq j<k\leq N_i}{\rm sgn}(x_{k}-x_{j})$ is the ``antisymmetric unit function''. 
Since $\mathcal{A}$ 
does not include time explicitly, it is also valid when the trap Hamiltonian is altered, and the time-dependent density profile resulting from 
this change can be calculated as \cite{GW00b}
%
$\rho(x,t)= N_i\!\!\int\vert\psi(x,x_{2},\dots,x_{N_i};t)\vert^{2} \d x_{2} \cdots\d x_{N_i}
=\sum_{n=1}^{N_i}\vert\varphi_{n}(x,t)\vert^{2}. \label{11}$
%
By reducing the trap capacity some of the $N_i$ atoms initially confined
may escape and $N$ will remain trapped.  
To determine whether or not sub-Poissonian statistics or a Fock state are achieved 
in the reduced trap we need to calculate  
the atom-number fluctuations.
We proceed by characterizing the TG trapped state by means of its variance
$\sigma_{N}^{2}=\la N^2(t)\ra-\la N(t)\ra^{2}$.
First note the general relation
\beqa
\la \widehat{n}(x)\widehat{n}(x')\ra_t\!\!&=&\!\!\la:\widehat{n}(x)\widehat{n}(x'):\ra_t
+\delta(x-x')\la \widehat{n}(x)\ra_t,
\eeqa
where the number field operator is $\widehat{n}(x)=\Psi^{\dagger}(x)\Psi(x)$, $\Psi(x)$,
$\Psi^{\dagger}(x)$ are the annihilation and creation operators at point $x$, and $:\quad:$ 
denotes normal ordering. 
In particular, within the Tonks-Girardeau regime,
%
\beqa
\la \widehat{n}(x)\widehat{n}(x')\ra_t&=&D(x,x';t)+\delta(x-x')\rho(x,t),
\eeqa
with \cite{GWT01}
\beqa
D(x,x';t)&=&\!\!N_i(N_i-1)\!\!\int\!\prod_{i=3}^{N_i}\d x_{i}
\vert\psi(x,x',x_3,\cdots,x_{N_i};t)\vert^2
\nonumber\\
&=&\rho(x,t)\rho(x',t)-|\Delta_{N_i}(x,x',t)|^2, 
\eeqa
and 
\beqa
\Delta_{N_i}(x,x';t)=\sum_{n=1}^{N_i}\varphi_{n}(x,t)^{*}\varphi_{n}(x',t).
\eeqa
The mean value of the number of particles within the trap and of its 
square can be obtained by integrating over $x,x'$,
\beqa
\la N(t)\ra&=&\int_{0}^{L_+}\d x\rho(x,t),
\nonumber\\
\la N^2(t)\ra&=&\int_{0}^{L_+}\int_{0}^{L_+}\d x \d x'\la\widehat{n}(x)
\widehat{n}(x')\ra_t,
\eeqa
where 
$L_+\sim L+\xi$ is large enough to include the bound-state tails 
in coordinate space. (For the trap configuration of Fig. 1 each bound state has a  penetration length $\{\xi_j\}$ beyond the well width $L$, 
so $\xi={\rm max}_j\xi_j$.) 
The atom number variance reads finally
\beqa
\sigma_{N}^2(t)
&=&\la N(t)\ra-\int_{0}^{L_+}\!\!\int_{0}^{L_+}\d x\d x'\Delta_{N_i}(x,x',t).
\eeqa
From it one can infer Poissonian statistics if $\sigma_{N}^2/\la N\ra\ge1$ and 
sub-Poissonian as long as $\sigma_{N}^2/\la N\ra<1$.
For a Fock state $\sigma_{N}=0$. 
\section{The sudden approximation} 
The requirement of adiabaticity, i.e., of a slow trap change, 
is a handicap which 
one would like to overcome. 
Achieving good fidelity with respect to the desired Fock state 
may require exceedingly long times, a fact that is even 
more critical whenever the interactions are finite, this is, 
for the Lieb-Liniger gas in which the splitting between adjacent levels (Bethe roots)
diminishes. It is thus useful to examine the opposite limit corresponding to a sudden 
trap change. For the TG gas, we shall find general and exact results 
which are a useful guide since small deviations from the sudden limit 
increasing the switching time may only 
improve the fidelity. 

In what follows we shall thus discuss  
the preparation of Fock states by an abrupt change of the trap potential
to reduce its capacity. 
Even though 
the arguments and results of this section are rather general, consider for concreteness 
the simple square trap configuration of Fig. 1, with an infinite wall on one side
and a flat potential (zero potential energy) on the other side,
\beqa
\label{trap}
V_{\alpha}(x)=
\left\{\begin{array}{l l l}
\infty &,\;\; x\leq 0\\
-V_{\alpha} &,\;\; 0<x\leq L_{\alpha} \\
0 &,\;\; x>L_{\alpha}
\end{array}\right.,
\eeqa
where the subindex $\alpha=i,f$ refers to the initial and final configuration respectively.
The corresponding eigenvalue problem is solved in the appendix, 
where both bound and scattering states are described. 
At $t=0$ the trap with shape $V_i(x)$, which holds an unknown number of particles $N_i$ (lower or equal than the initial capacity $C_i$)
is suddenly modified to the final trap $V_f(x)$, which supports  $C_f$ bound
states $\varphi_j^{f}(x)$, $j=1,...,C_f$, of energy $E_{j}^f<0$.
The process may consist on ``weakening'' the trap ($V_f<V_i$, $L_f=L_i$), 
``squeezing'' it ($L_f<L_i$, $V_f=V_i$), or a combination of the two ($V_f<V_i$ and $L_f<L_i$). 
%
%

The continuum part of the spectrum of the one-particle Hamiltonian with potential 
$V_f(x)$ is spanned by the scattering states $\chi_{k}^{f}(x)$, labeled by the 
incident wavenumber $k$. 
It follows from standard scattering theory \cite{Taylor},  
using the Riemann-Lebesgue lemma \cite{BH86}, that the contribution of 
the continuum states in the trap region $(x<L_+)$ vanishes asymptotically as $t\rightarrow\infty$, 
\beqa
\varphi_{n}(x,t)&=&\sum_{j=1}^{C_f}\la x\vert\varphi_j^f\ra
\la\varphi_j^f\vert\varphi_{n}^i\ra e^{-iE_{j}^{f}t/\hbar}\nonumber\\
& &+\int_{0}^{\infty} \d k\la x\vert\chi_k^f\ra
\la\chi_k^f\vert\varphi_{n}^i\ra e^{-i\frac{\hbar k^{2}t}{2m}}\nonumber\\
&\sim&\sum_{j=1}^{C_f}\la x\vert\varphi_j^f\ra
\la\varphi_j^f\vert\varphi_{n}^i\ra e^{-iE_{j}^{f}t/\hbar},
\eeqa
so that the dynamics in the trap is finally governed by the discrete part of the spectrum 
$\{\varphi_j^f|j=1,\dots,C_f\}$.
Therefore, asymptotically, the mean number and variance of trapped
atoms are 
\beqa
\la N(\infty)\ra&=&
\sum_{n=1}^{N_i}\la\varphi_n^i|\widehat{\Lambda}_f|\varphi_n^i\ra=\sum_{j=1}^{C_f}
\la \varphi_j^f|\widehat{\Lambda}_i|\varphi_j^f\ra,
\eeqa
note that $\la N(\infty)\ra\le C_f$, and  
\beqa
\sigma_{N}^2(\infty)&=&\la N(\infty)\ra-\sum_{n,m=1}^{N_i}\vert\la\varphi_m^i|\widehat{\Lambda}_f|\varphi_n^i\ra\vert^2
\nonumber\\
&=&\sum_j^{C_f} \la \varphi_j^f|(\widehat{\Lambda}_i-\widehat{\Lambda}_i\widehat{\Lambda}_f\widehat{\Lambda}_i)
|\varphi_j^f\ra,
\label{theory}
\eeqa
where 
\beqa
\widehat{\Lambda}_f=\sum_{j=1}^{C_f}|\varphi_j^f\ra\la\varphi_j^f|
\eeqa 
is the projector onto the
final bound states and 
\beqa
\widehat{\Lambda}_i=\sum_{n=1}^{N_i}|\varphi_n^i\ra\la\varphi_n^i|
\eeqa 
the projector onto the bound states occupied by the initial state. 
We may thus conclude that trap reduction 
can actually lead to the creation of Fock states with $\la N(\infty)\ra=C_f$ and $\sigma^2_N(\infty)=0$ when the initial states span 
the final ones, 
\beqa
\label{cond}
\widehat{\Lambda}_f\subset\widehat{\Lambda}_i.
\eeqa
%
%
A time scale for the validity of the asymptotic regime, after the trap switch, 
is provided by the lifetime of the lowest resonance of the final trap. 
A simple semiclassical estimate is $\tau=L_f(m/2V_f)^{1/2}$, 
assuming the escape of a classical particle 
from the well and approximating the resonance kinetic energy
by the potential depth.    
To prepare a $^{23}$Na Fock state in final trap of with $L_f\sim 50$ $\mu$m, supporting $N_f=10$ bound states, 
the asymptotic regime is approached for $t>\tau\sim 30$ ms. We insist though, that any slower potential change
will play in favour of the fidelity of the resulting Fock state
until the time scale in which losses and decoherence 
begin to play a role. 
\section{Trap weakening} 
A good guidance for Fock state preparation is provided by Eq. (\ref{theory}),
which leads to the requirement for $\widehat{\Lambda}_i$
to be an extension of $\widehat{\Lambda}_f$. 
This result is model-independent, 
and in particular 
it holds irrespective of the smoothness of the trapping potential.  
For illustration purposes we shall consider the square trap in Eq. (\ref{trap})
shown in Fig. \ref{pot} \cite{pons}.

\begin{figure}
\includegraphics[width=7.5cm,angle=0]{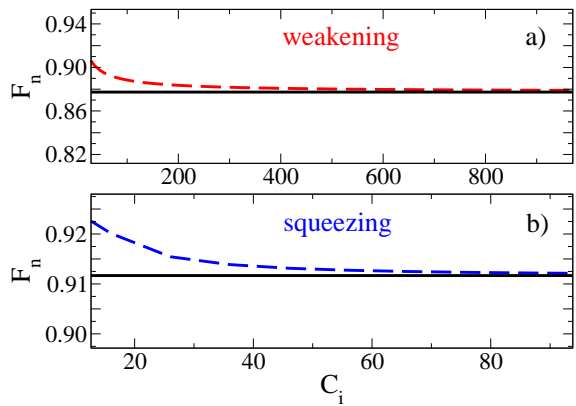}
\caption{\label{projector}
(color online) Limited efficiency on pure squeezing and weakening techniques. For a completely loaded initial trap ($N_i=C_i$) of depth $U_i=C_i^2\pi^2$, the resolution of a given bound state of the final trap, 
$\varphi_{n}^{f}$ ($n=5$, $U_f=25 \pi^2$), is quantified by the measure $F_n=\la\varphi_n^{f}|\hat{\Lambda}_i|\varphi_n^f\ra$, and shown to be limited in a) pure squeezing, and  b) pure weakening. For increasing capacity of the initial trap, $C_i$, 
$F_n$ tends to the probability to find the particle in the well region $[0,L_i]$ (weakening case, solid line, see Eq. (\ref{px})), or to the probability to find the particle in a momentum space region (squeezing case, 
solid line, see Eq.(\ref{kprojector})). 
} 
\end{figure}

In a trap-weakening scheme the potential is made shallower by reducing the depth 
from an initial value $V_i$ to $V_f$ (while $L_i=L_f$, see Fig. \ref{pot}), 
a procedure which has been successfully implemented 
%
to prepare states  
with sub-Poissonian statistics  
\cite{MSHCR05}.  
Though, in practice, the value of $N_{i}$ (and $C_f$) cannot be arbitrarily large  
because the Tonks-Girardeau regime requires a linear density $n\sim 1$ $\mu$m$^{-1}$, 
it is useful to consider a large number of particles in an initial box-like trap, $N_i\rightarrow\infty$, 
for which $\widehat{\Lambda}_{i}$ becomes the projector in the interval $[0,L_i]$,
\beq
\label{px}
\widehat{\Lambda}_{i}\sim\chi_{[0,L_i]}(\widehat{x}).
\eeq 
This asymptotic behavior is depicted in Fig. \ref{projector}a. 
Preparation of ideal atom number 
states by trap weakening is thus hindered by the suppression of the coordinate space tails 
which leak beyond the well along a given penetration length $\xi_j$ 
for each $\varphi_j^f$ \cite{note}.
%

%
%
%
%
%
%
\section{Trap squeezing} 
There is a simple alternative to trap weakening 
to achieve high-fidelity Fock states: atom-trap squeezing. 
Starting with a state of an unknown number of particles $N_i$, 
the trap width is reduced from an initial value $L_i$ to $L_f$ keeping the depth constant ($V_i=V_f$), 
as shown in Fig. \ref{pot}
(middle panel).  
The final trap supports $C_f$ 
bound states so that the excess of atoms is squeezed out of the trap. 
From a comparison of initial and final energy levels,
it is clear that a minimum number of initial particles is required for trap squeezing to work. 
Using for an estimate the levels of the infinite well we get $N_i>C_f L_i/L_f$.  
Trap squeezing works optimally for initial traps filled with atoms to the brim 
but it is not robust against partial filling. It is also less sensitive to threshold 
effects than trap weakening, in particular for low atom numbers.  

For a wide, filled initial trap, 
\beqa
\widehat{\Lambda}_i\sim\int_0^{\kappa}\d k|k^+\ra\la k^+|
\label{kprojector}
\eeqa
as $L\to\infty$, where $\la x|k^+\ra=\sqrt{2/\pi}\sin kx$, $x\geq 0$,  
satisfying $\la k^+|k'^+\ra=\delta(k-k')$ and the cut-off is at $\kappa\simeq U_{i}^{1/2}/L_i$, in terms of 
the dimensionless parameter $U_i=2mL_i^2V_i/\hbar^2$ 
($U_f$ is defined similarly in terms of the final values). This is illustrated in Fig. \ref{projector}b.
Therefore, trap squeezing may limit the fidelity of the final Fock state preparation due 
to the truncation of the tails ``in momentum space'', in the sense of Eq. (\ref{kprojector}), 
for $k>U_{i}^{1/2}/L_i$.
\section{Combined weakening and squeezing} 
%
%
%
%
\begin{figure}
\includegraphics[width=7.5cm,angle=0]{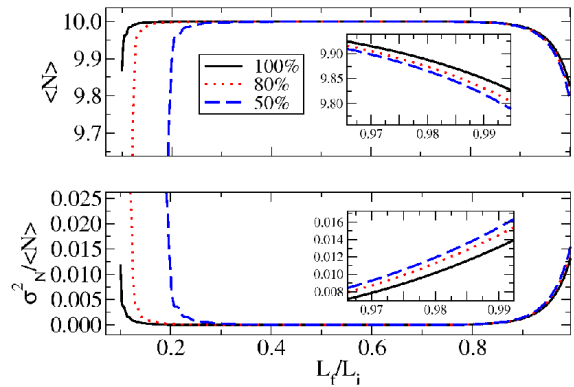}

\caption{\label{filling}
(color online) Asymptotic mean value (up) and variance (down) of the atom number distribution 
of a Tonks-Girardeau gas obtained by sudden trap reduction as 
a function of the width ratio between the final and initial trap. The initial trap with  $U_i/\pi^2=10^4$ supports a maximum of $C_i=100$ bound states, while  
the final configuration $U_f/\pi^2=10^2$ is limited to $C_f=10$. Different 
filling factors are considered for the initial trap. The left and right edges 
of each curve correspond to pure squeezing and weakening respectively, 
the later being remarkably less sensitive to the trap filling as shown in the inset. 
Any other point combines weakening and squeezing.} 
\end{figure}
%
%
\begin{figure}
\includegraphics[width=7.5cm,angle=0]{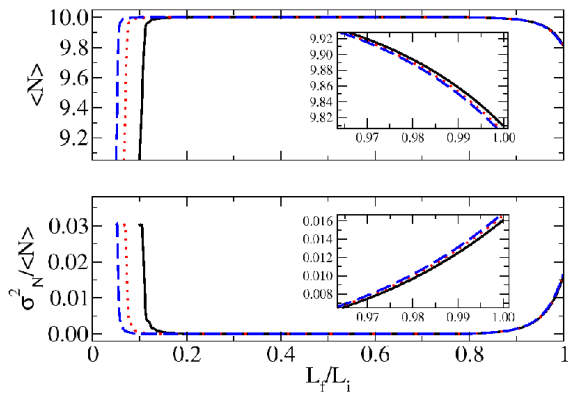}
\caption{\label{figu}
(color online) Asymptotic mean value (up) and variance (down) of the atom number distribution 
of a Tonks-Girardeau gas obtained by a sudden change of the trap potential 
for  $U_i/\pi^2=10^4$ (solid line), $2.25\times10^4$ (dotted line) and $4\times10^4$ (dashed line) 
keeping a constant initial filling factor of $90\%$. $U_f/\pi^2=10^2$ with $C_f=10$. The high-efficiency region 
increases with $U_i$ reducing the critical ratio between the final  and initial trap widths $L_{f}^c/L_i$.
} 
\end{figure}
%
From the previous discussion it follows that the optimal potential trap change to fullfill $\widehat{\Lambda}_f\subset\widehat{\Lambda}_i$ is a combination of weakening and squeezing. Let us choose two different families of isospectral traps characterized by $U_i$ and $U_f$, supporting $C_i$ and $C_f$ bound states respectively (In general $C_i>N_i$ because of partial filling of the initial trap, the filling factor being 
the ratio $N_i/C_i$). 
The energy level of the highest occupied state measured from the bottom of the trap, analogous to the Fermi level in the dual system, is denoted by $\varepsilon_i$, which depends on the filling factor of the $U_i$-trap,  while for the ideal final Fock state $\varepsilon_f\simeq V_f$.  

From an arbitrary potential $(L_i,V_i)$ in the $U_i$-family, the efficiency of the weakening-squeezing combination 
leading to a potential of the $U_f$ family varies with the filling factor and with 
the ratio of widths in the final and initial trap $L_f/L_i$. Figure \ref{filling} shows both the mean atom number and variance of the final states for different ratios and preparation states. Below a critical final width $L_{f}^c=(U_f/U_i)^{1/2}L_i$, the physical final depth $V_f$ is larger than the initial one $V_i$ and we disregard this possibility since the efficiency is very poor as expected from the failure of the condition (\ref{cond}) for $\varepsilon_f>\varepsilon_i$. 
The ratio $L_f/L_i=1$ is the limit of pure trap weakening, and $L_{f}^c/L_i$ that of pure trap squeezing; the limited efficiency of both extreme cases can be noticed in Fig. \ref{filling}. 
However, the combined process achieves pure Fock states and is robust with respect to different fillings in the initial trap and for a wide
range of final configurations. 
Moreover, the range of final configurations for which high-fidelity 
states are obtained increases with 
the initial potential $U_i$ keeping the filling constant as shown in Fig. \ref{figu}. 
Hence, a recipe to create a Fock state $|C_f\ra$ in a trap of width $L_f$ and depth $V_f$ will be as follows: 
choose the width of the initial trap broad enough in the sense $L_i\simeq L_f+r \xi$ ($r\gtrsim1$), where 
$\xi$ is the penetration length of the state $\varphi_{C_f}^f$ in the final trap.
Then, make sure that $\varepsilon_i> \varepsilon_f$, in such a way that the final state  
is contained in the initial subspace both in momentum and coordinate space. 
A deep and broad initial trap, with respect to the final one, 
provides in summary a safe starting point to create a Fock state
by sudden (or otherwise) trap reduction.

\section{Discussion and conclusion} 

In this work we have studied and compared strategies for atom Fock state creation in Tonks-Girardeau regime. 
We have shown that Fock states can be prepared even under a sudden 
trap-potential change. 
Thanks to the analysis of the atom number variance, we have determined that the key condition for Fock state creation is that the initial occupied bound states span the space of the final ones. This holds 
regardless of the trap shape details, and in particular does not depend on the 
potential-trap model. A combination of trap weakening and squeezing allows to resolve the ideal Fock state 
both in momentum and coordinate space. We close by noting that the Tonks-Girardeau regime is optimal with 
respect to the strength of interactions. In this regime, the three-body correlation function 
$g_3(\gamma)$ tends to vanish and therefore the losses of atoms from the trap 
by inelastic collisions are negligible \cite{GS03}. For gases with finite interactions in tight-waveguides, when the Lieb-Liniger model holds,  the quasimomenta obtained as solution of the Bethe equations \cite{Gaudin71,Cazalilla02,BGOL05,chinos}, are closer to each other for weaker interatomic interactions, whence it follows that the required time scale for the dynamics to be adiabatic is even larger than in the Tonks-Girardeau regime. Large spacings in the Bethe roots also imply that less precision 
is required in the control of $V_f$. Given that the interatomic interactions can be tuned through the Feschbach resonance technique, one can optimize the atom Fock state preparation by putting the system within a strong interaction regime in a first stage, followed by the controlled reduction of the potential trap (weakening and squeezing), and finally turning off slowly the interaction.

\begin{acknowledgments}
The authors acknowledge comments and discussions by
M. G. Raizen, H. Kelkar, and I. L. Egusquiza.  
This work has been supported by Ministerio de Educaci\'on y Ciencia (BFM2003-01003), and UPV-EHU (00039.310-15968/2004). A. C. acknowledges financial support by the Basque Government (BFI04.479).
\end{acknowledgments}

\appendix
\section{}

In this appendix we describe the spectrum of the Hamiltonian for the potential 
in Eq. (\ref{trap}) considered in the numerical examples.
Both the initial and final traps have the same functional dependence. 
Here we consider the general case for a trap of width $L$ and depth $V$ (dropping the index $\alpha=i,f$ for compactness), 
whose spectrum can be easily determined using matching conditions in the wavefunction
and its derivatives. 
The trap supports a finite set of $C$ bound states (its capacity) 
\beqa
\varphi_j(x)=
\mathcal{N}_j
\left\{\begin{array}{l l}
\sin(q_j x) &,\;\; 0\leq x< L \\[.25cm]
\sin(q_j L)e^{-\kappa_j(x-L)} &,\;\; x\geq L
\end{array}\right.
\eeqa
with $j=1,\dots,C$ and normalization constant
\beqa
\mathcal{N}_j=\left(\frac{L}{2}-\frac{\sin(2q_jL)}{4q_j}+\frac{\sin^2(q_jL)}{2\kappa_j}\right)^{-1/2}.
\eeqa
The eigenvalues are $E_j=(\hbar q_j)^2/2m-V<0$ where $\{q_j\}$ satisfy the trascendental equation
%
$\kappa_j=-q_j\cot q_jL$,
%
with $\kappa_j=(2mV/\hbar^2-q_j^2)^{1/2}>0$. 
The scattering states have the form 
\beqa
\chi_k(x)=\frac{1}{\sqrt{2\pi}}
\left\{\begin{array}{l l}
A \sin qx &,\;\; 0<x\leq L \\[.25cm]
e^{-ikx}-S(k)e^{ikx} &,\;\; x>L
\end{array}\right.,
\eeqa
where $q=\sqrt{k^2+2mV/\hbar^2}$ and the coefficient 
$A$ and the scattering matrix $S(k)$ are determined by imposing the usual matching conditions,
\beqa
A&=&-\frac{2ike^{-ikL}}{q\cos qL-ik\sin qL},\nonumber\\
S(k)&=&e^{-2ikL}\frac{q\cos qL+ik\sin qL}{q\cos qL-ik\sin qL}.
\eeqa
%



\begin{thebibliography}{10}
\bibitem{pons} The effect of the shape of the potential will be discussed elsewhere.
\end{thebibliography}
\end{document}